\documentclass[twocolumn,showpacs]{revtex4}

\usepackage{graphicx}%

\newcommand{\bb}{\begin{equation}}
\newcommand{\en}{\end{equation}}
\begin{document}

\title{Entropic Interactions in Suspensions of Semi-Flexible Rods:
Short-Range Effects of Flexibility}
\author{A.W.C. Lau, Keng-Hui Lin,}
\author{A.G. Yodh}
\affiliation{Department of Physics and Astronomy, University of Pennsylvania, Philadelphia, PA 19104}%

\date{\today}

\begin{abstract}
We compute the entropic interactions between two colloidal spheres
immersed in a dilute suspension of semi-flexible rods.  Our model
treats the semi-flexible rod as a bent rod at fixed angle, set by
the rod contour and persistence lengths. The entropic forces
arising from this additional rotational degree of freedom are
captured quantitatively by the model, and account for observations
at short range in a recent experiment.  Global fits to the
interaction potential data suggest the persistence length of
fd-virus is about two to three times smaller than the
commonly used value of $2.2\,\mu \mbox{m}$.
\end{abstract}
\maketitle

Colloidal dispersions exhibit a fascinating range of equilibrium and
non-equilibrium structures, and they have important impact
on our daily lives \cite{colloid}.  The interactions between
suspension constituents determines the stability of the dispersion against
flocculation, and the phase behavior of the colloid.  Quantitative models
and measurements of these interactions test our basic understanding about these
systems, and enable experimenters to better control suspension
behaviors and properties.  In this paper we focus on a particular class
of entropic interaction, exploring the forces between spheres in a
suspension of rodlike particles.  This system class has produced a
variety of interesting phases \cite{phase1,phase2,phase3},
and has stimulated several theoretical models \cite{karem1,theory2,theory3,theory4,theory5}
and a measurement \cite{lin} of the rod-induced depletion interaction.

The depletion attraction between two spheres immersed in a dilute suspension of
thin rods of length, $L_c$, was first considered by Asakara and Oosawa \cite{ossawa}.
Their most important physical insight was that rods in suspension
gain both translational and rotational entropy when the sphere surfaces come
within $L_c$ of one another.  Subsequent theories computed the attraction more
accurately within the Derjaguin approximation \cite{theory2,theory3}
and beyond \cite{karem1}.  However, in many practical scenarios
the rods are not rigid, and current theories do not account for the flexibility of the rods.
Indeed, flexibility effects can be important as evidenced by a recent interaction
measurement \cite{lin} of micron diameter spheres in suspensions of fd-virus; in this
case systematic deviations between experiment and ``rigid-rod'' theories were
found at short-range, and were suggested to arise as a result of the flexibility
of the fd-virus.  Flexible or bent rods have an additional degree of freedom:
the rotation about their central axis.  As the spheres get closer, this degree of freedom is
depleted, the system entropy increases, and the sphere interactions become even more
attractive.

\begin{figure}[bp]
\resizebox{2.1in}{2in}{\rotatebox{0}{\includegraphics{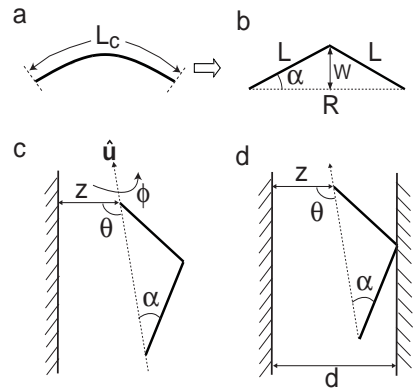}}}
\caption{(a) Typical configurations of a semi-flexible rod whose contour length $L_c$
is comparable to the persistence length $\ell_p$.  (b) Our approximation of
semi-flexible rod in (a); two stiff rods of length $L$ attached at a
fix angle.  (c) A bent rod near a flat wall and (d) confined between two walls.}
\label{figure1}
\end{figure}

A quantitative model for this observation is still lacking, and indeed a
complete theory of semi-flexible rods near surfaces remains a difficult task.
In this paper, we introduce a simple model to compute the depletion potential
between two spheres in a dilute solution of semi-flexible rods. We use the
model to quantitatively explain the experiments of Ref. \cite{lin}.  The model
accounts for the entropic effects of flexibility at short-range, and provides an
accurate fit of the measured interaction potentials.  The model also provides a
means to extract the persistence length $\ell_p$ and the contour length $L_c$ of
the suspended semi-flexible rods from interaction potential data.  Global fitting
of the data suggests that the persistence length of fd-virus is
two to three times smaller than the commonly used value of $2.2\,\mu\mbox{m}$ \cite{persistence}.

Our model relies on the assumption that if the rods are sufficiently stiff, they
may be accurately approximated by two rods of length $L= L_c/2$, attached together at a fixed angle
$\pi - 2 \alpha$, as shown in Fig. \ref{figure1}.  The angle $\alpha$ may be estimated by
$\alpha = \cos^{-1} R/ L_c $, where $R \equiv \left ({\langle\, {\bf R}^2 \rangle}\right )^{1/ 2}$
is the average end-to-end distance.  $R$ is related to $\ell_p$ and $L_c$ by \cite{doi}\bb
\langle \,{\bf R}^2 \rangle = 2 L_c \ell_p + 2 \ell_p^2 \left ( e^{- L_c /\ell_p} -1 \right ).
\label{endtoend}
\en
This approach simplifies the problem, while still capturing the essential physics.
In particular, we show that the part of the depletion potential associated with
new rotational degrees of freedom is short ranged, {\em i.e.} of order the bent rod width,
$W = L \sin \alpha$.  Importantly, when $W$ is significantly less than the particle diameter,
the rotational part of the depletion interaction can be treated within the Derjaguin
approximation \cite{colloid}.

\begin{figure}[bp]
\resizebox{2in}{1.2in}{\rotatebox{0}{\includegraphics{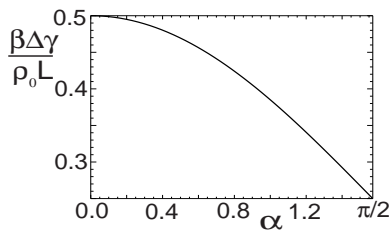}}}
\caption{The surface tension of a bent rod in the presence of a
flat plane wall.} \label{figure2}
\end{figure}

\begin{figure}[bp]
{\par\centering
\resizebox*{3in}{3.5in}{\rotatebox{0}{\includegraphics{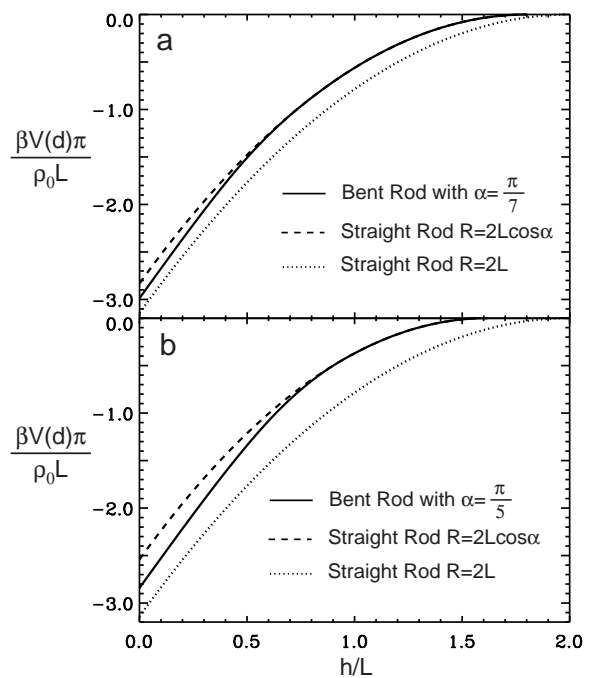}}}
\par}
\caption{The depletion interaction $V(d)$ [Eq. (\ref{potential})]
between two planar walls (the solid curve) mediated by a bent rod of contour length
$L_c = 2 L$ with (a) $\alpha = \pi/7$ and (b) $\pi/5$. The dashed
curve is the depletion interaction of a straight rod with
$R = 2 L \cos \alpha$. At large distances, they show little
difference but the restriction on the additional degree of freedom
at shorter distances gives rise to a stronger attraction in $V(d)$, which is
bounded below by the potential of a straight rod of $R = 2 L$ (the dotted line).  This
is qualitatively the effect observed in experiment of Ref. \cite{lin} (See also Fig. \ref{figure4}).}
\label{figure3}
\end{figure}

In the presence of repulsive walls (see Fig. \ref{figure1}),
the rotational degrees of freedom of a bent
rod are restricted.  Consider a bent rod with one end displaced by $z$ from
the wall and with orientation $(\hat{{\bf u}}, \phi)$.
The probability of finding a rod in such a configuration is
given by the Boltzmann factor: $ f({\bf r}, \hat{{\bf u}}, \phi)
\propto \exp[ - \beta U_{e}({\bf r}, \hat{{\bf u}}, \phi)]$, where
$\beta = 1/k_B T$, $k_B$ is the Boltzmann constant, and $T$ is the temperature.  The
hard wall potential $U_{e}$ is infinite if any part of the rod
touches the wall and is otherwise zero.  We consider the case where
the concentration of the rods is sufficiently low so that the thermodynamics are
well characterized by the Grand potential of an ideal gas of rods\bb
\Omega = - N k_B T \int d^3 {\bf r} \int d^2 {\bf u} \int d \phi \, f({\bf r}, {\bf u}, \phi).
\en
Here $N$ is the number of rods. We define the surface tension by the difference\begin{eqnarray}
& & \Delta \gamma = {\Omega - \Omega_0 \over S} \nonumber \\
&& =\rho_0 k_B T \int {d^3 {\bf r} \over S}  \int {d^2 {\bf u} \over 4 \pi} \int {d \phi \over 2 \pi}
\left [ 1 - e^{- \beta U_{e}({\bf r}, {\bf u}, \phi)} \right ].
\label{gamma}
\end{eqnarray}
Here $\rho_0$ is the average density of the rods and $S$ is the surface area of the wall.
To compute the integral in Eq. (\ref{gamma}), we enumerate all the configurations
of the bent rod just touching the walls.

Let us first consider a single flat wall,
as shown in Fig. \ref{figure1}.  There are three regions to consider:
(i) $0 < \alpha < \pi/4$, (ii) $ \pi/4 < \alpha < \pi /3$, and (iii) $\pi/3 < \alpha < \pi/2$.
For $\alpha < {\pi / 4}$, we observe that when $\theta_1(z,\alpha)< \theta < \theta_2(z,\alpha)$,
for\begin{eqnarray}
\theta_1(z,\alpha) & =& \cos^{-1} \left [ {z \over 2 L \cos \alpha } \right ] \\
\theta_2(z,\alpha) &=&  \alpha + \cos^{-1} z/L,
\end{eqnarray}
the rotation of the rod about its symmetry axis is restricted
to $\phi_a < \phi < 2 \pi - \phi_a$, where\bb
\phi_a(z,\theta)= \cos^{-1} \left[ {z \over L \sin \alpha \sin \theta }
- \cot \theta \cot \alpha \right ].
\en
Using this construction, the surface tension is\bb
{\beta \Delta \gamma(\alpha) \over \rho_0}= {L \cos \alpha \over 2}
+ {L \over 2 \pi} \int_0^{\sin 2 \alpha} dx \int_{ \theta_1}^{\theta_2}d \theta \sin \theta \,\phi_a(x,\theta),
\en
where $x= z/L$.  Similarly, for ${\pi \over 2} > \alpha  > {\pi \over 4}$, we have
\begin{eqnarray}
{\beta \Delta \gamma(\alpha) \over \rho_0} &=& {L \cos \alpha \over 2} +
{L \over 2 \pi} \int_0^{\sin 2 \alpha} dx \int_{ \theta_1 }^{\theta_2}d \theta \sin \theta \,\phi_a(x,\theta) \nonumber \\
&+& {L \over 2 \pi} \int_{\sin 2 \alpha}^{1} dx
\int_{\alpha -\cos^{-1} x }^{\theta_2} d \theta \sin \theta \, \phi_a(x,\theta).
\label{onewall}
\end{eqnarray}
In Fig. \ref{figure2}, we plot $\Delta \gamma(\alpha)$ as a function of $\alpha$.
Note that the limiting values,
$\Delta \gamma(0)  = {1 \over 2} \rho_0 k_B T  L$ and $\Delta \gamma(\pi/2) ={1 \over 4} \rho_0 k_B T L$,
agree with previous results \cite{karem2}.

\begin{figure}[bp]
\resizebox{3in}{2.2in}{\rotatebox{0}{\includegraphics{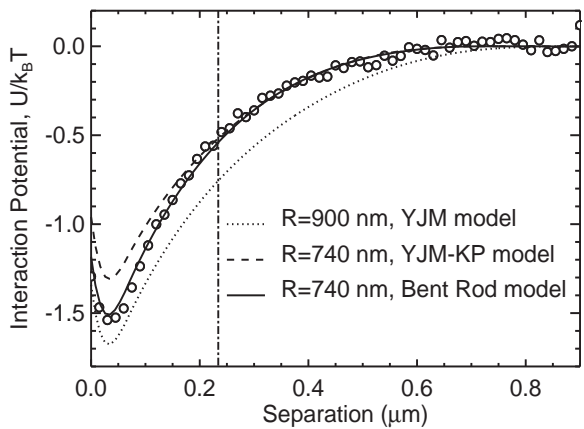}}}
\caption{Interaction potential between a pair of $1.0\,\mu$m silica spheres in a suspension of
fd virus with concentration $0.67\,\mbox{mg/ml}$.  The dotted (dashed) lines are generated
by the YJM model with $R = L_c = 900$ nm  ($R = 740$ nm; $L_c = 900$ nm).  The solid lines are
generated by Eq. (\ref{sphere}) with $R=740$ nm and $L_c = 900$ nm.  Clearly,
the agreement of experimental data and our model which includes the addition
rotational degree of freedom of a bent rod is excellent. The dash-dotted vertical
line indicates $W = 0.23\,\mu$m.}
\label{figure4}
\end{figure}

We now turn to the calculation for two walls (see Fig. \ref{figure1}).
Since the rods are stiff and $\alpha$ is small in the experiment of interest,
we focus on the case where $\alpha < \pi/4$. For a given separation of the walls $d$,
we divide the interval $0<z<d/2$ into different
regions, wherein $\theta_1, \theta_2, \theta_3,$ and $\theta_4$ take on different values.
The new angles are\begin{eqnarray}
\theta_3(z,\alpha) & =& \pi - \alpha - \cos^{-1} \left [ {d - z \over 2 L  } \right ] \\
\theta_4(z,\alpha) &=&  \pi - \cos^{-1} \left [ { d- z \over 2L \cos\alpha }\right].
\end{eqnarray}
If $\theta_1 < \theta < \theta_2$ and $\theta_2 < \theta_3$,
$\phi$ is restricted to $\phi_a < \phi < 2\pi - \phi_a$.
If $\theta_3 < \theta < \theta_4$, $\phi$ is restricted to
$ 0 < \phi < \pi - \phi_b$ and $ \pi + \phi_b < \phi < 2 \pi$ with\bb
\phi_b(z,\theta)= \cos^{-1}
\left [ 1 - { L \cos( \pi - \theta- \alpha) - (d-z) \over L \sin \alpha \sin (\pi -\theta) }\right ].
\en
When $\theta_2 > \theta_3$, $\phi$ is further restricted to $\phi_a < \phi < \pi - \phi_b$ and
$\pi+\phi_b < \phi < 2 \pi - \phi_a$ if $\theta_3 < \theta < \theta_2$.
Thus, the depletion potential per unit area defined by
$V(d) = \Delta\gamma[d,\alpha] - \Delta\gamma[\infty,\alpha]$
is\bb
V(d) = - \rho_0  k_B T
\left [ L \cos \alpha \left ( 1 - {d \over 2 L \cos \alpha } \right )^2  + \Gamma(d,\alpha) \right ],
\label{potential}
\en
where\begin{eqnarray}
\Gamma(d,\alpha) &=& {1 \over \pi} \int_0^{L \sin 2 \alpha}
dz \int_{ \theta_1}^{\theta_2}d \theta \sin \theta \,\phi_a(z,\theta;\alpha)\nonumber \\
&-& {1 \over \pi} \int' dz \int' d\theta \sin\theta\,\phi_a(z,\theta;\alpha)\nonumber \\
&-& {1 \over \pi} \int' dz \int' d\theta \sin\theta\,\phi_b(z,\theta;\alpha).
\end{eqnarray}
Here $'$ indicates integrations over phase space restricted
to the allowed values. Fig. \ref{figure3} depicts the depletion
potential between two walls for different $\alpha$.  At large distances the potential
is determined by the ``end-to-end'' distance $R$ of the rod. At short distances the
rotational degree of freedom becomes important and increases
the attraction between walls.  The potential is bounded below
by the potential of a straight rod with length $2L$.
Although our calculation has been done for two walls,
we expect the same qualitative features to hold for two spheres.

\begin{figure}[bp]
\resizebox{3in}{4in}{\rotatebox{0}{\includegraphics{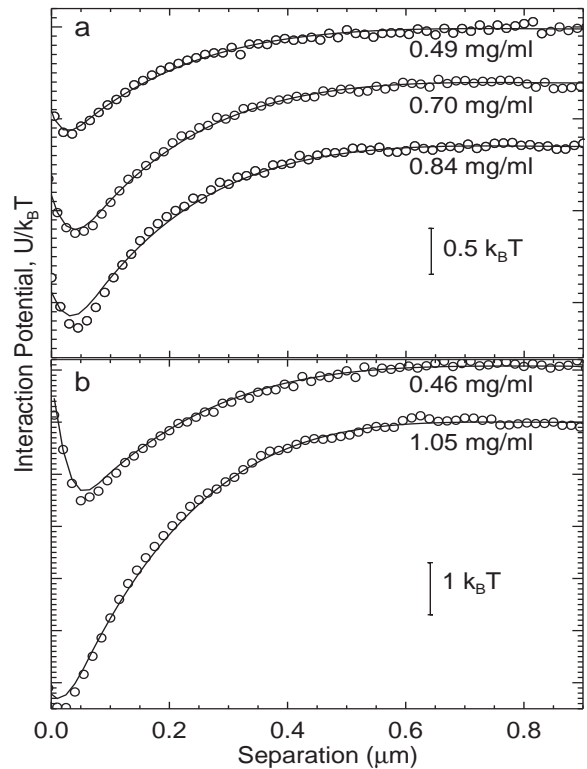}}}
\caption{Interaction potential between pairs of (a) $1.0\,\mu$m and (b) $1.6\,\mu$m
silica spheres in a suspension of fd virus with different concentration.
The solid lines are generated by Eq. (\ref{sphere}), with best fit parameters that
give smallest $\chi^2$.}
\label{figure5}
\end{figure}

$V(d)$ can be written as a sum of 2 pieces.
The first term is the depletion potential of a {\em straight} rod with
length $R = 2 L \cos \alpha$ \cite{karem2}.  The second term depends only on the additional
rotational degree of freedom of the bent rod.  Moreover, the range
of $\Gamma(d,\alpha)$ is of order of the width of the bent rod, $W = L \sin \alpha$, which
is small compared to the sphere radius in Ref. \cite{lin}. These observations
suggest that to approximate the depletion potential for two spheres,
the latter term may be treated in the Derjaguin approximation,
while the first term replaced by the YJM rigid-rod model \cite{karem1}.
Thus, we write\bb
\beta U_s(h) = - \rho_0 a R^2
\left [ K \left ({ h\over R};{ a\over R}\right )
+ {\pi \over R^2} \int_{h}^{\infty}dx\,\Gamma(x,\alpha) \right ],
\label{sphere}
\en
where $a$ is the sphere radius and $h$ their closest surface separation.
$K(h/R;a/R)$ is the potential between two spheres due to a straight rod
of length $R$, which reduces to the Derjaguin expression  $K_D(h/R) ={\pi \over 6} ( 1 - h/R)^3$
in the limit $a/R \gg 1$ \cite{karem1}.

Fig. \ref{figure4} displays a typical experimental data set of
Ref. \cite{lin} with three different models -- (i)
the YJM model (dotted line), whose potential is given by the first term in Eq. (\ref{sphere})
with $R = 900\,\mbox{nm}$, the contour length of fd, (ii) the YJM-KP model (dashed line),
whose potential is given by the first term in Eq. (\ref{sphere})
with $R = 740\,\mbox{nm}$, and (iii) the bent rod model (solid line), $U_s(h)$ in
Eq. (\ref{sphere}) with $R= 740\,\mbox{nm}$ and $L_c =900\,\mbox{nm}$.
The circles are experimental data for $1.0\,\mu\mbox{m}$ diameter silica particle
in a dilute ($0.67\,\mbox{mg/ml}$) solution of fd virus.  The theory curves are computed
with no free parameters and are then numerically blurred to account for the instrument's spatial resolution
(see Ref. \cite{lin}).  Clearly, our model gives the best fit to the experimental data.
In particular, both YJM and YJM-KP models, while having approximately the right magnitude and shape,
fail to account for the overall curvatures of the experimental curve.  Further,
while the YJM-KP model agrees with most of the data at large $h$,
our model clearly accounts for the depth of the measured potential near contact.

In order to explore the best fits more quantitatively,
we computed the $\chi^2$ value of our models for all data sets with $R$ ranging
from $720 - 825\,\,\mbox{nm}$ and $L_c=880, \,900$, and $920 \,\mbox{nm}$.
If a fixed concentration (measured experimentally) is assumed, $\chi^2$ is smallest
for $R=780 \,\mbox{nm}$ and $L_c=920 \,\mbox{nm}$. If the concentration is allowed to vary
within its $\pm 5 \%$ experimental error, then $\chi^2$ is smallest
for $R=740 \,\mbox{nm}$ and $L_c = 900 \, \mbox{nm}$.
Fig. \ref{figure5} shows best fits for each of concentration.
Note that the width $W \sim 200\,\mbox{nm}$, is smaller
than the radius of the colloidal spheres. This justifies {\em a posteriori} the Derjaguin approximation
made in Eq. (\ref{sphere}).  Furthermore, we can estimate $\ell_p$ using Eq. (\ref{endtoend})
and the values for $R$ and $L_c$ above, yielding $\ell_p \simeq 850\,\mbox{nm}$ (fixed concentration)
and $\ell_p \simeq 680\,\mbox{nm}$ (variable concentration).  Our results for $L_c$ of
fd are consistent with the literature, {\em i.e.} $850\,\mbox{nm} < L_c < 920\,\mbox{nm}$ \cite{contour}.
However, our values for $\ell_p$ should be contrasted to the often-quoted value,
$\ell_p = 2.2\,\mu\mbox{m}$ \cite{dls,song}.  The latter is based on a fitting of dynamic
light scattering data with theoretical models \cite{song,theory}, whose assumptions may
well be questioned in the light of our results.  Indeed, smaller values of $\ell_p$
have also been reported based on dynamic structure factor models
of semi-flexible filaments \cite{augustin}, and using electron microscopy \cite{jaytang}.

We have presented a simple analytical model for the depletion
interaction between two spheres mediated by semi-flexible rods, and demonstrated its
quantitative agreement with experimental data.  Our theoretical model combined
with interaction measurements provides a basis for extracting the
persistence length of a semi-flexible rod.

\begin{acknowledgments}
We thank R.D. Kamien, Tom Lubensky, C.M. Marques,
and Fyl Pincus for valuable discussions. A.W.C.L. is particularly
grateful to K. Yaman for sharing his insights.  This work
is supported by the NIH Grant HL67286, and partially supported by the NSF through
DMR-0203378 and the MRSEC Grant DMR 00-79909.
\end{acknowledgments}

\end{document}